\documentclass[reprint,cha]{revtex4-1}

\usepackage{graphicx,setspace}
\usepackage{dcolumn}
\usepackage{bm}
\usepackage{hyperref}
\usepackage{natbib}

\begin{document}

\title{Stochastic and Vibrational Resonances in a Uni-junction Transistor Relaxation Oscillator}%
\author{Ajit Mahata}
\email{ajitnonlinear@gmail.com}
\affiliation{Department of Physics, National Institute of Technology
  Sikkim, Ravangla, Sikkim-737139, India.}
	
\author{Utpal Deka}
\email{udeka1rediffmail.com}
\affiliation{Department of Physics, Sikkim Manipal Institute of Technology, Majitar, Sikkim-737139, India.}

\author{Md.Nurujjaman}
\email{jaman\_nonlinear@yahoo.co.in}
\affiliation{Department of Physics, National Institute of Technology
  Sikkim, Ravangla, Sikkim-737139, India.}

\begin{abstract}
 The effects of perturbation of a weak periodic signal (WPS) assisted by noise or a high frequency signal (HFS) respectively on an excitable uni-junction transistor relaxation oscillator (UJT-RO) is presented here. When the perturbation by a WPS modulated with the noise is optimum, the UJT-RO has been observed to produce regular dynamics that mimics the WPS maximum, which is termed as stochastic resonance (SR). Interestingly, when the noise component is replaced by a HFS, the system shows the same SR kind of behavior, which is called vibrational resonance (VR). Here the system produces the dynamics that mimics the WPS resembling the SR assisted by the optimum level of HFS. Both SR and VR have been confirmed through  PSPICE simulation. The results show that the regular dynamics that mimics the WPS in case of HFS modulation is better than the noise perturbation. 
\end{abstract}

\maketitle
\section{Introduction}
\label{section:intro}
The study of the noise-induced phenomena in an excitable~\cite{PRE:nurujjaman2009,yang2012vibrational} or bistable~\cite{casado2004effects,fauve1983stochastic} systems has received considerable attention, as there is a possibility to detect low level signals in such system with the help of noise perturbation~\cite{PRE:nurujjaman2008,PRE:nurujjaman2010}. The counter-intuitive view of improving the response of a system to an external force has several manifestations. It is observed that when such perturbation is purely noise, it gives rise to coherence resonance (CR)~\cite{PRE:nurujjaman2009,pikovsky1997coherence,giacomelli2000experimental,kiss2003experiments}. It is well known that perturbation enforced by a weak periodic signal (WPS) modulated with noise produces stochastic resonance (SR)~\cite{benzi1981mechanism,benzi1982stochastic,nicolis1981stochastic}. Such phenomena have brought about a major change in the outlook of the effect of noise in several scientific field~\cite{bulsara1993stochastic}. The main characteristic of a SR is that the system response to a WPS reaches a maximum at an optimal level of noise~\cite{hanggi2002stochastic,jung1993periodically,moss1994stochastic}. Confirmation of SR has been demonstrated in various nonlinear systems theoretically~\cite{gammaitoni1998stochastic,durrant2011suprathreshold,burada2008entropic} as well as experimentally~\cite{mcnamara1988observation,samardak2009noise,tiwari2016intrinsic}. The SR like phenomena can also be reproduced with a high frequency signal (HFS) perturbation substituting the noise. The phenomenon of improvement of system dynamics, induced by the optimum level of a HFS in presence of a WPS, is called vibrational resonance (VR)~\cite{landa2000vibrational}. VR has been demonstrated in various system both theoretically~\cite{casado2004effects} and experimentally~\cite{stan2009stochastic,yao2010signal,baltanas2003experimental,chizhevsky2003experimental}. 
 
Most of the noise or HFS-invoked phenomena can be observed in the excitable systems under suitable parametric conditions~\cite{PRE:nurujjaman2009,yang2012vibrational,Chasolfrac:bordet2013}. The main characteristic of an excitable system is that it has an internal threshold depending on certain important parameter(s), and the threshold must be crossed in response to an external perturbation to excite the system dynamics. The excitability is usually achieved by setting the parameter(s) so that the system remains close to the threshold. External perturbation by noise at excitable state produces the dynamics that mimics the characteristic of the system is called CR. When the system is perturbed by a WPS along with noise, it produces SR. On a similar note, VR can be obtained, when the noise is replaced by HFS. In both SR and VR phenomena, the WPS helps the system to cross the threshold periodically to obtain periodic oscillatory or spiky dynamics that mimics the WPS~\cite{pre:duan2014}.

\begin{figure*}[bt]
\centering
\textsc{}\includegraphics[angle=0, width=16.0cm, height=9.85 cm]{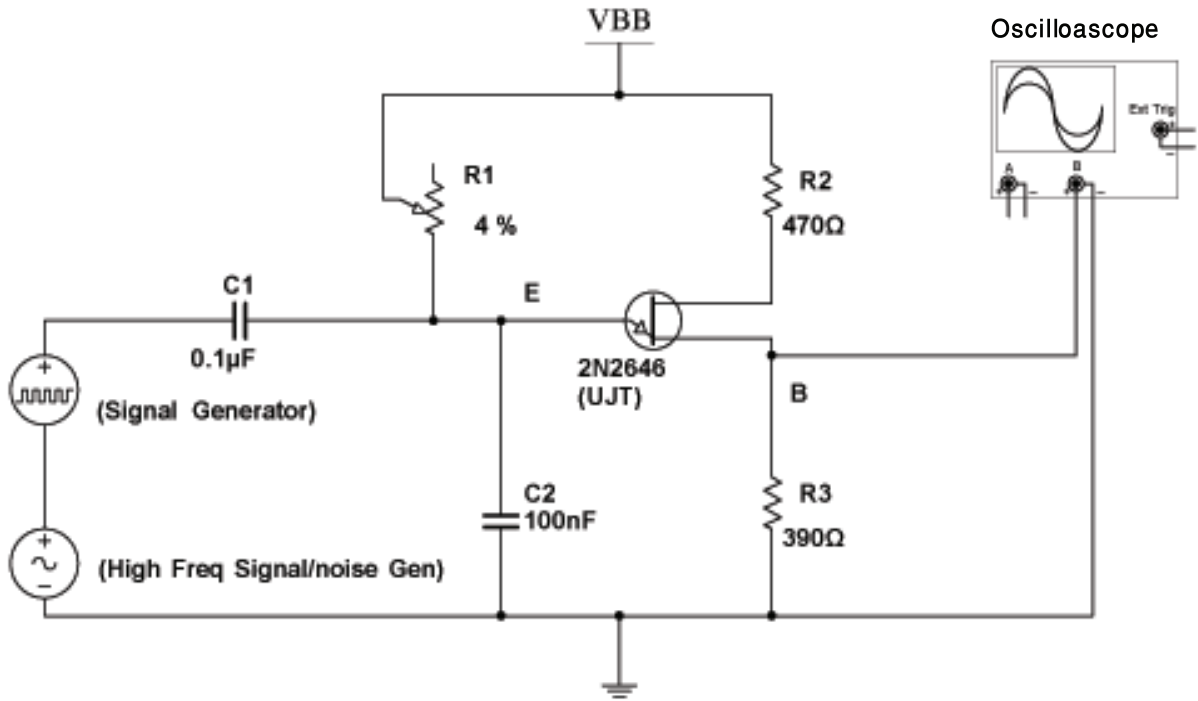}
\caption{Circuit Description of the UJT Relaxation Oscillator}
\label{fig:ckt}
\end{figure*}
In this paper, we present the experimental evidence and characterization of the phenomena of the SR and VR in a uni-junction transistor relaxation oscillator (UJT-RO)~\cite{PRE:nurujjaman2009,PRE:nurujjaman2010}. These phenomena have also been reproduced by PSPICE simulation using multisim, and the simulation results are very much in agreement with experimental results. The UJT-RO is a well-known excitable system, which can be used to study the noise-invoked phenomena~\cite{PRE:nurujjaman2009,PRE:nurujjaman2010}. It is observed that excitability in a UJT-RO can be obtained by setting base to emitter voltage, $V_{BE}=3.2~V$. The detailed experimental conditions have been discuss in Sec.~\ref{section:setup}.

In this work we have experimentally demonstrated the existance of both SR and VR in UJT-RO and the same is confirmed by PSPICE simulation using multisim. Rest of the paper has been organized as follows: The experiment and autonomous dynamics is presented in Sec.~\ref{section:setup}, experimental and simulation results have been presented in Sec.~\ref{section:result}, and finally a conclusion has been drawn in Sec.~\ref{section:conclusion}.

\section{Experimental setup and autonomous dynamics}
\label{section:setup}

Experiments have been performed to demonstrate SR and VR using UJT-RO, which contains 2N2646-UJT. The WPS and HFS used in the experiments have been generated by using 33500B Agilent signal generators, which are connected to UJT-RO through a $0.1~\mu$F capacitor to the circuit as shown in Fig.~\ref{fig:ckt}. The response of the system, which is the voltage across the base ($V_BO$), has been recorded by a DPO 3054 Tektronix digital oscilloscope. We have chosen the following set of values for capacitors and resistors in our experiments: $R_1=100~k\Omega$ (variable resistance), $R_2=470~\Omega$, $R_3=390~\Omega$, $C_1=0.1~\mu$F, $C_2=0.1~\mu$F, with a tolerance of 5\% for all the components as shown in Fig.~\ref{fig:ckt}. In the present experiments, the UJT-RO was operated at $V_{BB}=5V$. Depending on the voltage ($V_E$), periodic oscillation were observed, and the output at the point B ($V_{B}O$) was recorded in the oscilloscope. Here, $V_E$ acts as a control parameter, which was varied by using a variable resistance ($R_1$). The detailed of the autonomous dynamics of UJT-RO has been presented in Ref.~\cite{PRE:nurujjaman2009}.

The frequency and amplitude of the WPS were considered to be $f=100~Hz$ and 100 mV respectively. The bandwidth of the noise was set at 200 kHz. Frequency of the HFS was kept at $200~kHz$, which is much higher than the frequency of the WPS as well as that of the characteristic frequency of the system. Frequency of the HFS was chosen such that there is no possibility of the system getting excited by the HFS perturbation alone. A square pulse of width, 10 ns was used as a WPS for the VR experiment. The modulated amplitude of the noise and HFS acted as a perturbation on the control parameter for both SR and VR experiments. The base voltage was chosen in such a way that the system remains at its excitable state after application of the WPS in both cases. On application of certain level of noise or HFS, system started to show the desired dynamics. The detailed experimental results obtained from the experiments have been presented in the next Section [Sec.~\ref{section:result}].

\section{RESULTS AND DISCUSSIONS}
\label{section:result}

Stochastic resonance (SR) and vibrational resonance (VR) phenomena are observed experimentally in a UJT-RO experiment. The effects of noise and HFS on the non-linear system have been presented below.
			
\subsection{Effect of noise on nonlinear system: Stochastic resonance}
\label{subsection:sr}
\begin{figure*}[ht]
\centering
\includegraphics[width=16.0cm, height=9.85 cm]{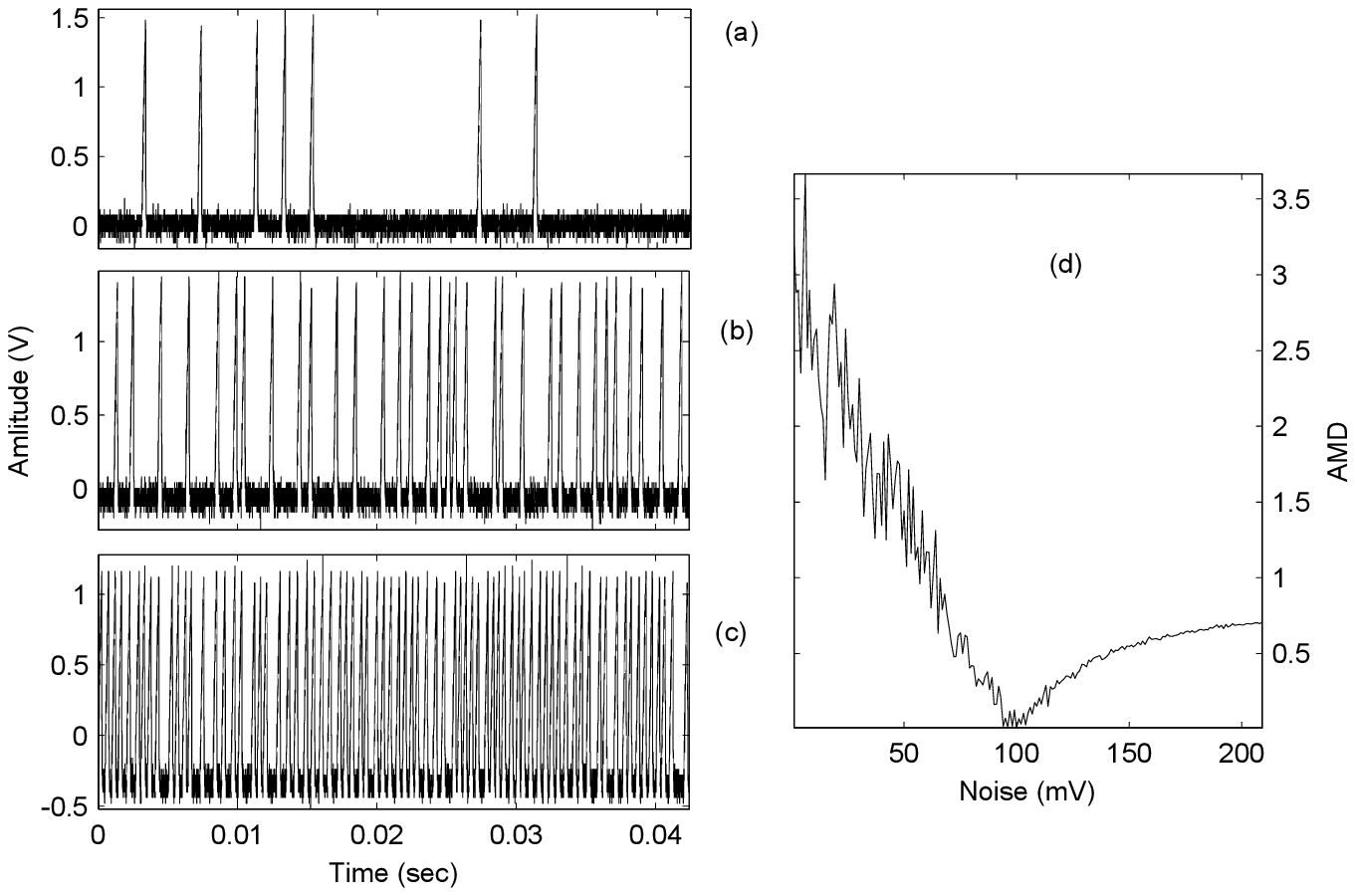}
\caption{Left panel: Time series obtained from of the UJT-RO across the base resistance to show the stochastic resonance. (a) low (48mV) (b) Optimum (100mV) (c) high level (150mV) noise. Right panel: AMD versus noise graph}
\label{fig:SR}
\end{figure*}
	
Figs.~\ref{fig:SR}(a)-~\ref{fig:SR}(c) show the time series of the output ($V_B $O), taken from the base terminal (B) of the UJT-RO for different noise levels. Fig.~\ref{fig:SR}(d) represents the absolute mean difference (AMD) as a function of noise amplitude in mV. AMD has been used to quantify the information transferred from sub-threshold signal to system response. AMD is defined as, $ AMD=abs[mean(t_p/\delta-1)]$, where $t_p$ is the inter-peak interval of the response signal and $\delta$ is the mean peak interval of the sub-threshold signal. We have applied a WPS of amplitude 20 mV to the emitter terminal as shown in Fig.~\ref{fig:ckt}. The value of AMD=1.5 as shown in the Fig.~\ref{fig:SR}(d) corresponds to noise of amplitude 48 mV, and the corresponding time series is shown in Fig.~\ref{fig:SR}(a). In such noise value, the activation threshold is seldom crossed, and a random spike sequence is generated as shown in Fig.~\ref{fig:SR}(a). As we increase the noise amplitude, the value of AMD decreases [Fig.~\ref{fig:SR}(d)], and reaches a minimum (almost zero) at a noise level of 100 mV. The corresponding time series has been shown in Fig.~\ref{fig:SR}(b). The minimum of the AMD in Fig.~\ref{fig:SR}(d) shows that the response of the system is optimum around this region. It indicates a maximum regularity or uniformity has been achieved in the dynamics. The maximum detection of WPS is possible around the minimum of the AMD. Further increase in the noise intensity destroys the regularity of the time series, that is evident from the increase in value of AMD. Irregular time series at high noise level has been observed as shown in Fig.~\ref{fig:SR}(c). From Fig.~\ref{fig:SR}(c) it is clear that irregularity or noise dominates the system dynamics. The production of regular dynamics at optimum level of noise is termed as SR [minimum region in Fig.~\ref{fig:SR}(d)].
				
\begin{figure}[ht]
\centering
\includegraphics[width=8.6cm]{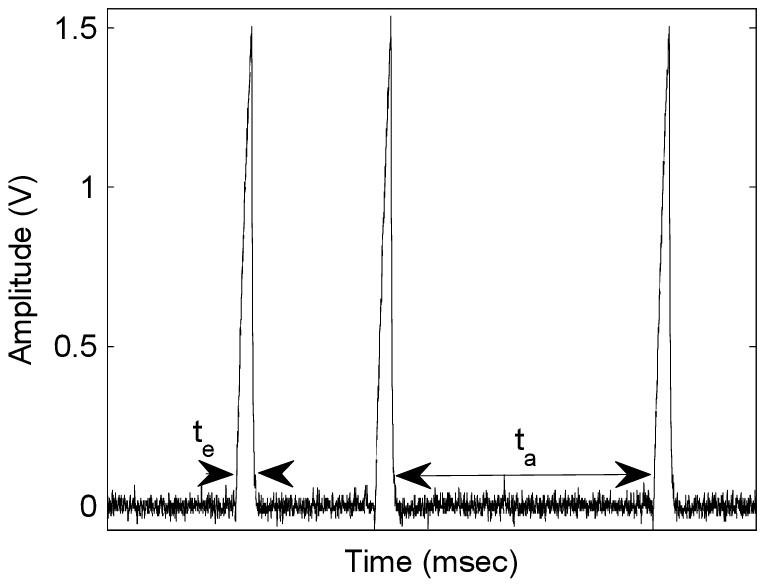}
\caption{}
\label{fig:excur}
\end{figure}
				
\begin{figure*}[ht]
\centering
\includegraphics[width=16.0cm, height=9.85 cm]{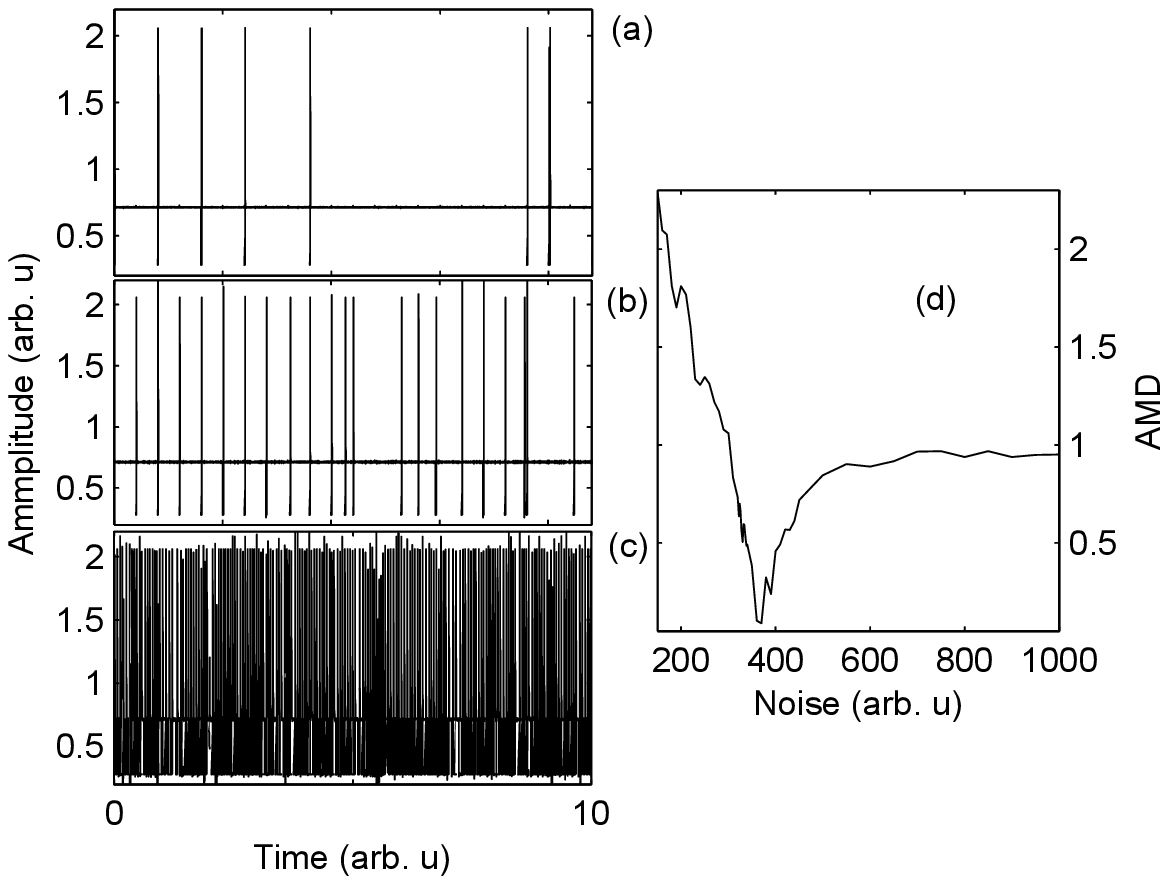}
\caption{PSPICE simulated SR recorded at base (V$_{B}$O):
The right panel shows the AMD  as a function of noise. Left panel: The time series of the output at (a) low 300 (arb.u),
(b) optimum 400 (arb.u) and (c) high-level 600 (arb.u) of noise.}
\label{fig:PSPICE SR}
\end{figure*}
Fig.~\ref{fig:SR}(d) shows that the system attains a state where coherency of oscillation  increases with the increase of noise intensity, and subsequently decreases at higher value of noise. At lower noise level, the inter peak distance or time is dominated by the activation time ($t_a$) of the limit cycle oscillation, which varies significantly. Again with the increase of noise level, the activation time ($t_a$) decreases and excursion time ($t_e$) dominates, and hence the system reaches regularity. The activation time is the time required to excite the system, where as the excursion time is the time required to return from the excited state to the fixed point. The Fig.~\ref{fig:excur} shows that the pulse duration is given by $t_p= t_a + t_e$. For small noise level, $t_a>>t_e$ and the period of the time series is dominated by activation time, and hence $t_p \approx t_a$. In the case for high level of noise, the activation time ($t_a$)  is negligible and the excursion time ($t_e$) dominates the dynamics, and hence $t_p \approx t_e$. At an optimum level of noise, the number of crossings of threshold is much larger than the excursion frequency of oscillation. The limit cycle oscillation is distorted at high frequency noise level, and hence the AMD increases, which has been shown in Fig.~\ref{fig:SR}(d).
			
We have also carried out PSPICE simulation using multisim to verify the experimental results of SR. The set point of the threshold was set at 3 V for the present experiment. For simulation, variable resistance $R_1$ was chosen slightly higher than the experimental value to obtain the relaxation oscillations. The weak periodic signal of amplitude 20 mV and noise are added through $0.1~\mu$F capacitor and data was taken at base terminal ($V_BO$). We have chosen the following set of values of capacitors and resistors for simulation of SR: $R_2=470~\Omega$, $R_3=390~\Omega$, $C_1=0.1~\mu$F and $C_2=100$ nF as shown in Fig.~\ref{fig:ckt}, and all of these components have tolerance of 5\%. Figs.~\ref{fig:PSPICE SR}(a)-~\ref{fig:PSPICE SR}(c) show the three time series of $V_BO$ at different noise level, and corresponding AMD has been shown in Fig.~\ref{fig:PSPICE SR}(d). The simulation results are in good agreement with our experimental results.			
	\subsection{Effects of high frequency signal: Vibrational resonance}
	\label{subsection:vr}
	
\begin{figure*}[ht]
\centering
\includegraphics[width=16.0cm,height=9.85 cm]{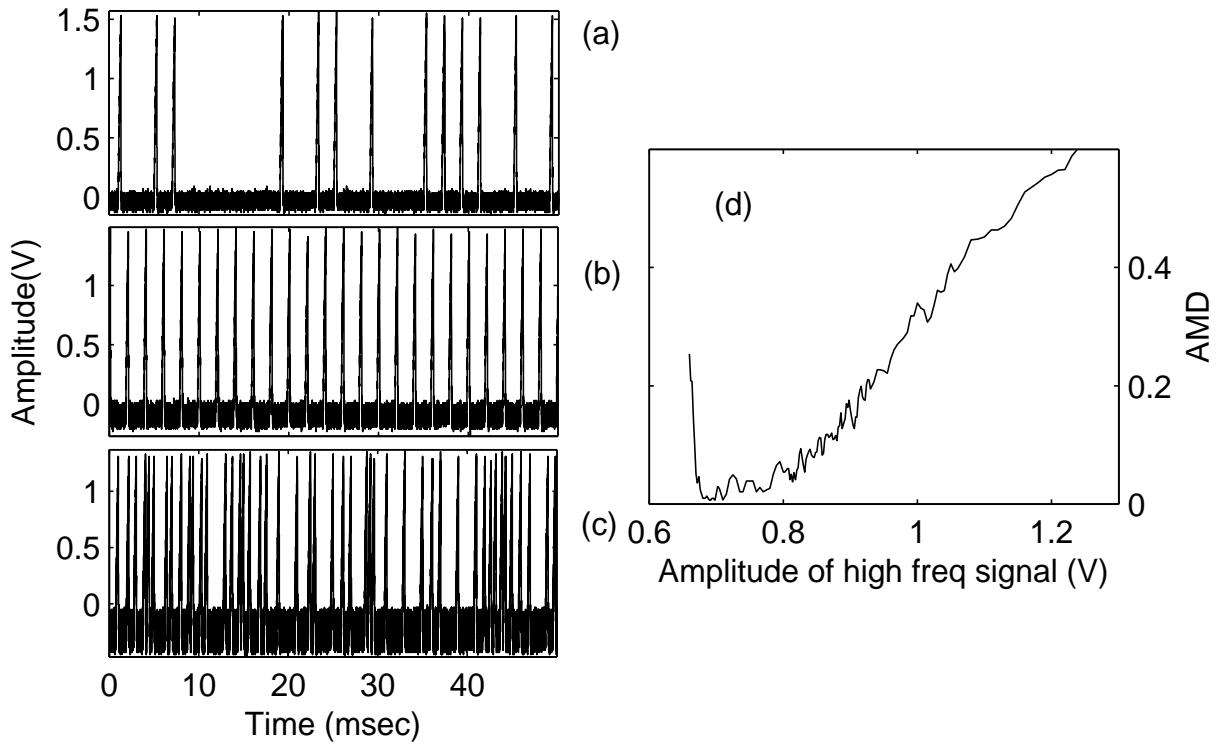}
\caption{Emergence of vibrational resonance for the output recorded at base (V$_{B}$O):
The right panel shows the AMD as a function of noise . Left panel: The time series of the output at (a) low (0.65V),
(b) optimum (0.7V) and (c) high-level (0.9V) of noise.}
\label{fig:VR}
\end{figure*}
	
\begin{figure*}[ht]
\centering
\includegraphics[width=16.0cm,height=9.85 cm]{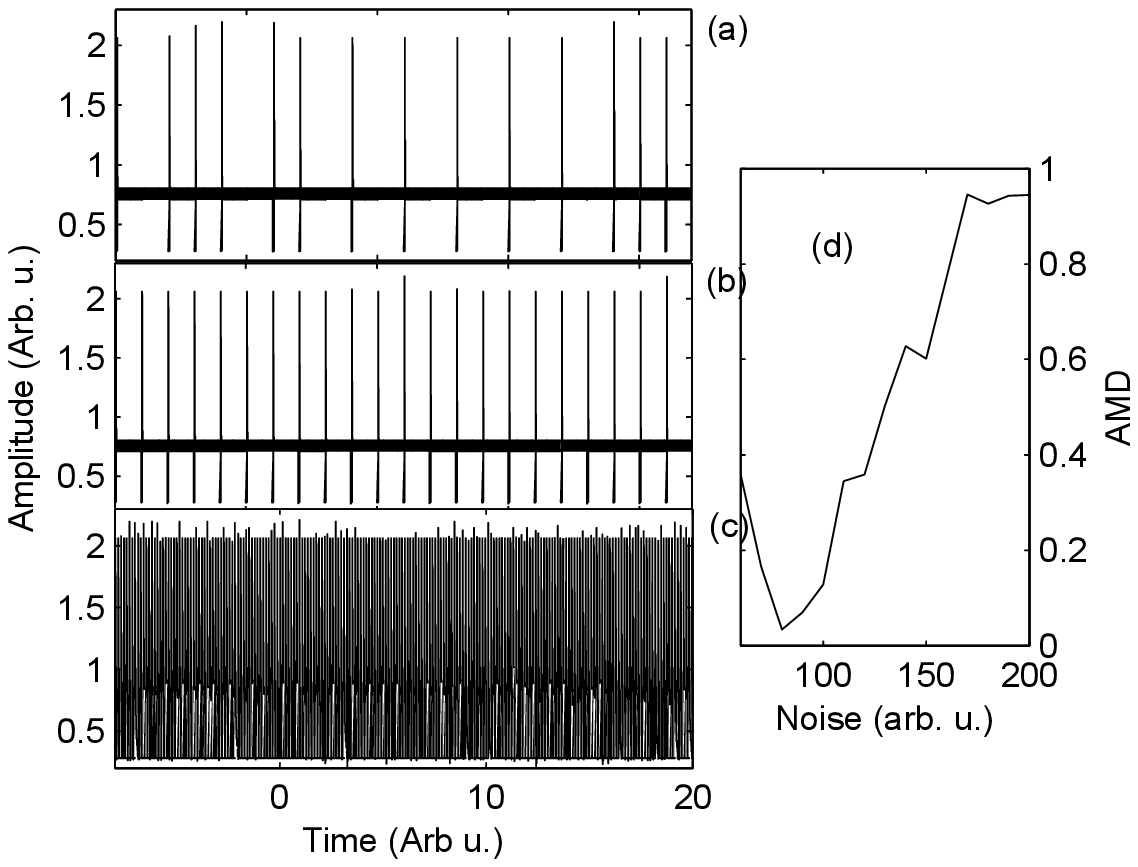}
\caption{PSPICE simulated VR recorded at base (V$_{B}$O):
The right panel shows the AMD as a function of noise. Left panel: The time series of the output at (a) low 25 (arb.u),
(b) optimum 50 (arb.u) and (c) high-level 125 (arb.u) of noise.}
\label{fig:PSPICE VR}
\end{figure*}

It is already seen in Subsec.~\ref{subsection:sr} that noise can play an important role in the dynamics of excitable non-linear system. Here we have studied the effect of HFS to produce SR kind of phenomena, which is generally termed as vibrational resonance (VR). When the UJT-RO was perturbed with a HFS of smaller amplitude in place of noise in SR experiment as described in Subsec.~\ref{subsection:sr}, same kind of random spikes have been observed in the time series as shown in Fig.~\ref{fig:VR}(a). The corresponding AMD value comes out to be high as shown in Fig.~\ref{fig:VR}(d). Further it is interesting to note that as we increase the amplitude of HFS, the value of AMD decreases, which is clearly visible from Fig.~\ref{fig:VR}(d). It reaches a minimum around a noise level of 0.7 V, and the corresponding time series around this point has been given in Fig.~\ref{fig:VR}(b). The minimum in the Fig.~\ref{fig:VR}(d) represents the optimum response of the system, i.e, a maximum regularity or uniformity is obtained in the dynamics. So we can conclude that around the minimum, optimal detection of WPS can be achieved. Further increase in the intensity of HFS, regularity of the time series gets destroyed due to HFS perturbation as shown in Fig.~\ref{fig:VR}(c), which is also clear from the AMD value [Fig.~\ref{fig:VR}(d)]. The higher value of AMD in Fig.~\ref{fig:VR}(d) represents a time series of Fig.~\ref{fig:VR}(c). It shows that, the AMD increases with irregularity in the dynamics. At higher amplitude of HFS perturbation, the system dynamics get affected only by HFS suppressing the effect of WPS. 
				
Therefore, coherency of the oscillation produced by the system is negligible when the amplitude of HFS is very low and  increases with the increase in the amplitude of HFS. At low level of HFS the inter peak distance or time is dominated by the activation time of the limit cycle oscillation, which varies randomly. With increases in the amplitude of HFS, the activation time decreases and hence the AMD also decreases. At an optimum level of HFS, crossing of threshold by the external perturbation, is much larger than the excursion frequency of oscillation, and hence we get optimum regular oscillation, which is also clear from the minimum of AMD graph [Fig.~\ref{fig:VR}(d)]. The limit cycle oscillation gets distorted at HFS, and hence AMD increases with the increases of HFS amplitude which is also clear from Fig.~\ref{fig:VR}(d).

				PSPICE simulation of VR was carried out under the similar conditions mentioned in Subsec.~\ref{subsection:sr}. The time series thus obtained are presented along with the AMD as shown in Fig.~\ref{fig:PSPICE VR}. The simulation results are almost in agreement with the experimental results.
\section{conclusion}
\label{section:conclusion}
   
   In conclusion, occurrence of Stochastic resonance (SR) and Vibrational resonance (VR) has been studied experimentally in UJT-RO experiment. The experimental results have also been verified using PSPICE simulation. In both SR and VR, system has been seen to mimic the low frequency weak periodic signal with optimum level of noise and high frequency signal respectively. Performance of VR has been found more effective than SR to produce the dynamics that mimics the weak perturbation. 
	We expect that these kind of experimental studies may help to detect weak signals using nonlinear dynamics that may be applicable in various scientific fields. Further, it also gives a unique method to design an excitable system, where implementation of external perturbation is easier. 

\section*{Acknowledgment}
We would like to acknowledge the help of Mr. S. Sankar, Electrical and Electronics Engineering Department, NIT Sikkim during the experiment. We would like to acknowledge the support of Director NIT, Sikkim. We also acknowledge the help of Mr. Krishna Sharma and Mr. Bhanu Chetrry of Sikkim Manipal Institute of Technology, Sikkim for helping in simulation.

\bibliographystyle{aipnum4-1}
\bibliography{ref}

\end{document}